\documentclass[conference,letterpaper]{IEEEtran}
\usepackage{makeidx}
\usepackage[british,UKenglish,USenglish,english,american]{babel}
\usepackage[utf8]{inputenc}
\usepackage[dvips]{graphicx}
\usepackage{graphicx,url}
\usepackage{amssymb}
\usepackage{psfrag,amsmath}
\usepackage{multirow}
\usepackage[Lenny]{fncychap}
\usepackage[dvipsnames]{xcolor}
\usepackage{fancyhdr}
\usepackage{makeidx}
\usepackage{epsfig}
\usepackage{graphicx,color}
\usepackage{subfig}
\usepackage{array}
\usepackage{arydshln}
\usepackage[all]{xy}
\usepackage{float}
\usepackage{hyperref}
\usepackage{multicol}
\usepackage[margin=10pt,font=scriptsize,labelfont=bf,labelsep=endash]{caption}

\usepackage[T1]{fontenc}
\usepackage{ifthen}
\usepackage{cite}
\usepackage{stfloats} 
\usepackage{blindtext} 

\newcommand{\Fn}{\mathbb{F}_2^n}

\newcommand{\bx}{\mathbf{x}}
\newcommand{\by}{\mathbf{y}}
\newcommand{\bz}{\mathbf{z}}
\newcommand{\bzero}{\mathbf{0}}

\newcommand{\be}{\mathbf{e}}

\newcommand{\rank}{\mathrm{rank}}
\newcommand{\Ran}{\mathrm{\mathbf{Rank}}}

\begin{document}

\title{Equivalence classes of small tilings of the Hamming cube}

\author{%
		\IEEEauthorblockN{Gabriella Akemi Miyamoto}
		\IEEEauthorblockA{Imecc - 
						University of Campinas\\
				gabriellaakemimiyamoto@gmail.com}}
\maketitle

\section{Introduction}
The study of tilings is a major problem in many mathematical instances, which is studied in two main different approaches: when considering the existence (or obstructions to the existence) of a tiling with a given tile and the other considering classification of tilings. Considering the Hamming cube $\mathbb{F}_2^n$, the small tilings, that is, tilings considering tiles with $8=2^3$ elements, were classified in \cite{vardy}. The authors list a total of $193$ different tiles. As the authors noted, many of those tiles can be obtained one from the other by a linear map. In this work, we are concerned with a particular class of linear maps, the class of permutations of coordinates. This is of interest since a permutation is an isometry of the Hamming cube, considering the Hamming metric. We show here that, up to an isometry, all those $193$ tiles can be reduced to $15$ classes. The proof is done by explicitly showing the permutation (represented in cycles) that identify each tile with a given representative.

As in \cite{vardy}, we separate the tiles by their rank, where the \textbf{rank of} a set $A\subset\Fn$ is the dimension of the vector subspace spanned by the vectors of $A$, ie, $\rank (A)=dim\langle A \rangle$; The \textbf{rank of a tiling} $(D,C)$ is $\rank (D,C)=\rank (D)$.

For each rank, $D$ is identified as $D=B_H(\bzero,1)\cup S$, where $B_H(\bzero, 1) \subset\mathbb{F}_2^{\rank(D)}$ is the Hamming ball of center $\bzero$ and radius 1 and  $S\subset \mathbb{F}_2^{\rank(D)}$ is a set such that $|S|=0,1,2$ or $3$. They stated some conditions over the weight of the vectors in $S$ in such way that $D$ is a tile. As an example, let $D=B_4(\bzero,1)\cup \{\bx,\by,\bz\}=\{\bzero, \be_1,\be_2,\be_3,\be_4,\bx,\by,\bz\}$, then if $\omega(\bx)=\omega(\by)=\omega(\bz)=3$, we have $D$ is a tile. Respecting the conditions over $\bx,\by$ and $\bz$, we have that \[\bx,\by,\bz\in \left\{ \begin{array}{c}
	\be_1+\be_2+\be_3  \\
	\be_1+\be_2+\be_4  \\
	\be_1+\be_3+\be_4  \\
    \be_2+\be_3+	\be_4 \\
\end{array}\right\}. 
\]

Then, all the possibilities for $D$ is 
\vspace{3pt}

{\color{OrangeRed}
$D_1=\left\{ \begin{array}{ccc}
\bzero &  & \be_4 \\
\be_1 &  & \be_1+\be_2+\be_3 \\
\be_2 &  & \be_1+\be_2+\be_4 \\
\be_3 &  & \be_1+\be_3+\be_4 \\
\end{array}\right\}$ }

\vspace{9pt}

$D_2=\left\{ \begin{array}{ccc}
\bzero &  & \be_4 \\
\be_1 &  & \be_2+\be_1+\be_3 \\
\be_2 &  & \be_2+\be_1+\be_4 \\
\be_3 &  & \be_2+\be_3+\be_4 \\
\end{array}\right\} \leftrightarrow {\color{OrangeRed}D_1}, P=(12)$

\vspace{9pt}

$D_3=\left\{ \begin{array}{ccc}
\bzero &  & \be_4 \\
\be_1 &  & \be_4+\be_2+\be_1 \\
\be_2 &  & \be_4+\be_3+\be_1 \\
\be_3 &  & \be_4+\be_3+\be_2 \\
\end{array}\right\}\leftrightarrow {\color{OrangeRed}D_1}, P=(14)$

\vspace{9pt}

$D_4=\left\{ \begin{array}{ccc}
\bzero &  & \be_4 \\
\be_1 &  & \be_3+\be_2+\be_1 \\
\be_2 &  & \be_3+\be_1+\be_4 \\
\be_3 &  & \be_3+\be_2+\be_4 \\
\end{array}\right\}\leftrightarrow {\color{OrangeRed}D_1}, P=(13)$

\vspace{9pt}

Notice that $D_2$, $D_3$ and $D_4$ are all equivalent to $D_1$. To verify this fact, consider the permutations $(12)$, $(14)$ and $(13)$, respectively. Therefore, the four cases can be reduced to one.

\vspace{6pt}

Throughout this paper, the representatives will be in colorful characters. If $B$ is equivalent to $\textcolor{RedOrange}{A}$, we put the symbol $B\leftrightarrow \textcolor{RedOrange}{A}, P=(ij)$ to indicate which tile $B$ is equivalent to and which is the permutation that leads $B$ to $\textcolor{RedOrange}{A}$. 

We denote $D_i^r$ the tiling with rank $r$ and counting index $i$. 

\vspace{3pt}  

Consider the tile $D$ with $\rank(D)=3$. For this case, we have just one possible tile. 

\vspace{6pt}

$\Ran (D)=3$.

\vspace{3pt}

We have $D=\{\bzero, \be_1, \be_2, \be_3, \be_1+\be_2, \be_1+\be_3,\be_2+\be_3,\be_1+\be_2+\be_3\}$. 

\vspace{6pt}

The next case is when $D$ has rank equal to 4. 

\vspace{3pt}

$\Ran (D)=4$.

\vspace{3pt}

Let $D=\{\bzero, \be_1,\be_2,\be_3,\be_4,\bx,\by,\bz\}$, where $\omega_H(\bx)=\omega_H(\by)=\omega_H(\bz)=2$ and $\omega_H(\bx+\by)=\omega_H(\bx+\bz)=\omega_H(\by+\bz)=2$, then all the possibilities for $D$ are \\

{\color{Aquamarine}	
${\color{Aquamarine}D_1^4}=\left\{ \begin{array}{cc}
	\bzero &   \be_4 \\
	\be_1 &   \be_1+\be_2 \\
	\be_2 &   \be_1+\be_3 \\
	\be_3 &   \be_1+\be_4 \\
\end{array}\right\} $}

\vspace{9pt}
 
{\color{CarnationPink}	
$D_2^4=\left\{ \begin{array}{cc}
\bzero &   \be_4 \\
\be_1 &  \be_1+\be_2 \\
\be_2 &  \be_1+\be_3 \\
\be_3 &   \be_2+\be_3 \\
\end{array}\right\}$}

\vspace{9pt}

$D_3^4=\left\{ \begin{array}{cc}
\bzero &   \be_4 \\
\be_1 &   \be_1+\be_2 \\
\be_2 &   \be_1+\be_4 \\
\be_3 &   \be_2+\be_4 \\
\end{array}\right\} \leftrightarrow {\color{CarnationPink}D_2^4}, P=(34)$

 \vspace{9pt} 
 
 $D_4^4=\left\{ \begin{array}{cc}
\bzero &   \be_4 \\
\be_1 &   \be_2+\be_1 \\
\be_2 &   \be_2+\be_3 \\
\be_3 &   \be_2+\be_4 \\
\end{array}\right\} \leftrightarrow {\color{Aquamarine}D_1^4}, P=(12)$

\vspace{9pt}

$D_5^4=\left\{ \begin{array}{cc}
\bzero &   \be_4 \\
\be_1 &   \be_1+\be_3 \\
\be_2 &   \be_1+\be_4 \\
\be_3 &   \be_3+\be_4 \\
\end{array}\right\} \leftrightarrow {\color{CarnationPink}D_2^4}, P=(24)$

\hspace{9pt} 

$D_6^4=\left\{ \begin{array}{cc}
\bzero &   \be_4 \\
\be_1 &  \be_3+\be_1 \\
\be_2 &   \be_3+\be_4 \\
\be_3 &   \be_3+\be_2 \\
\end{array}\right\} \leftrightarrow {\color{Aquamarine}D_1^4}, P=(13)$

\vspace{9pt}

$D_7^4=\left\{ \begin{array}{cc}
\bzero &   \be_4 \\
\be_1 &   \be_4+\be_1 \\
\be_2 &   \be_4+\be_3 \\
\be_3 &   \be_4+\be_2 \\
\end{array}\right\} \leftrightarrow {\color{Aquamarine}D_1^4}, P=(14)$

 \vspace{9pt}

$
D_8^4=\left\{ \begin{array}{cc}
\bzero &   \be_4 \\
\be_1 &   \be_3+\be_4 \\
\be_2 &   \be_3+\be_2 \\
\be_3 &   \be_2+\be_4 \\
\end{array}\right\} \leftrightarrow {\color{CarnationPink}D_2^4}, P=(143)$

\vspace{9pt}

Let $D=\{\bzero, \be_1,\be_2,\be_3,\be_4,\bx,\by,\bz\}$, where $\omega_H(\bx)=\omega_H(\by)=2$ and $\omega_H(\bz)=3$ and $\omega_H(\bx+\by)=2$ and $\omega_H(\bx+\bz)=\omega_H(\by+\bz)=1$. Then, all the possibilities for $D$ are \\

{\color{Peach}
${\color{Peach} D_9^4}=\left\{ \begin{array}{cc}
\bzero &   \be_4 \\
\be_1 &   \be_1+\be_2 \\
\be_2 &   \be_1+\be_3 \\
\be_3 &   \be_1+\be_2+\be_3 \\
\end{array}\right\} $}

\vspace{9pt}

$ D_{10}^4=\left\{ \begin{array}{cc}
\bzero &   \be_4 \\
\be_1 &   \be_1+\be_2 \\
\be_2 &   \be_1+\be_4 \\
\be_3 &   \be_1+\be_2+\be_4 \\
\end{array}\right\} \leftrightarrow {\color{Peach} D_9^4}, P=(34)$

\vspace{9pt}

$D_{11}^4=\left\{ \begin{array}{cc}
\bzero &   \be_4 \\
\be_1 &   \be_2+\be_1 \\
\be_2 &   \be_2+\be_3 \\
\be_3 &   \be_2+\be_1+\be_3 \\
\end{array}\right\} \leftrightarrow {\color{Peach} D_9^4}, P=(12)$

\vspace{9pt}

$ D_{12}^4=\left\{ \begin{array}{cc}
\bzero &   \be_4 \\
\be_1 &   \be_2+\be_1 \\
\be_2 &   \be_2+\be_4 \\
\be_3 &   \be_2+\be_1+\be_4 \\
\end{array}\right\} \leftrightarrow  {\color{Peach} D_9^4}, P=(12)(34)$

\vspace{9pt}

$D_{13}^4=\left\{ \begin{array}{cc}
\bzero &   \be_4 \\
\be_1 &   \be_1+\be_3 \\
\be_2 &   \be_1+\be_4 \\
\be_3 &   \be_1+\be_3+\be_4 \\
\end{array}\right\} \leftrightarrow  {\color{Peach} D_9^4},  P=(24)$
\vspace{9pt}

$D_{14}^4=\left\{ \begin{array}{cc}
\bzero &  \be_4 \\
\be_1 &   \be_3+\be_1 \\
\be_2 &   \be_3+\be_4 \\
\be_3 &   \be_3+\be_1+\be_4 \\
\end{array}\right\} \leftrightarrow  {\color{Peach} D_9^4}, P=(13)(24)$
\vspace{9pt}

$D_{15}^4=\left\{ \begin{array}{cc}
\bzero &  \be_4 \\
\be_1 &   \be_3+\be_1 \\
\be_2 &   \be_3+\be_2 \\
\be_3 &   \be_3+\be_2+\be_1 \\
\end{array}\right\} \leftrightarrow  {\color{Peach} D_9^4},  P=(13)$
\vspace{9pt}

$D_{16}^4=\left\{ \begin{array}{cc}
\bzero &   \be_4 \\
\be_1 &   \be_4+\be_1 \\
\be_2 &   \be_4+\be_3 \\
\be_3 &   \be_4+\be_3+\be_1 \\
\end{array}\right\} \leftrightarrow  {\color{Peach} D_9^4},  P=(124)$

\vspace{9pt}

$D_{17}^4=\left\{ \begin{array}{cc}
\bzero &   \be_4 \\
\be_1 &   \be_4+\be_1 \\
\be_2 &   \be_4+\be_2 \\
\be_3 &   \be_4+\be_2+\be_1 \\
\end{array}\right\} \leftrightarrow {\color{Peach} D_9^4},  P=(134)$

\vspace{9pt}

$D_{18}^4=\left\{ \begin{array}{cc}
\bzero &   \be_4 \\
\be_1 &  \be_3+\be_4 \\
\be_2 &  \be_3+\be_2 \\
\be_3 &  \be_3+\be_2+\be_4 \\
\end{array}\right\} \leftrightarrow {\color{Peach} D_9^4},  P=(143)$

\vspace{9pt}

$D_{19}^4=\left\{ \begin{array}{cc}
\bzero &   \be_4 \\
\be_1 &   \be_4+\be_3 \\
\be_2 &   \be_4+\be_2 \\
\be_3 &   \be_4+\be_3+\be_2 \\
\end{array}\right\} \leftrightarrow {\color{Peach} D_9^4},  P=(14)$

\vspace{9pt}

$D_{20}^4=\left\{ \begin{array}{cc}
\bzero &   \be_4 \\
\be_1 &   \be_2+\be_3 \\
\be_2 &   \be_2+\be_4 \\
\be_3 &   \be_2+\be_3+\be_4 \\
\end{array}\right\} \leftrightarrow {\color{Peach} D_9^4},  P=(142)$

\vspace{9pt}

Let $D=\{\bzero, \be_1,\be_2,\be_3,\be_4,\bx,\by,\bz\}$, where $\omega_H(\bx)=2$ and $\omega_H(\by)=\omega_H(\bz)=3$ and $\omega_H(\bx+\by)=1$ or $\omega_H(\bx+\bz)=1$. Then, all the possibilities for $D$ are \\

{\color{Periwinkle}
${\color{Periwinkle}
D_{21}^4}=\left\{ \begin{array}{ccc}
\bzero &  & \be_4 \\
\be_1 &  & \be_1+\be_2 \\
\be_2 &  & \be_1+\be_2+\be_3 \\
\be_3 &  & \be_1+\be_2+\be_4 \\
\end{array}\right\} $}

\vspace{9pt}

{\color{Salmon}
${\color{Salmon}
D_{22}^4}=\left\{ \begin{array}{ccc}
\bzero &  & \be_4 \\
\be_1 &  & \be_1+\be_3 \\
\be_2 &  & \be_1+\be_2+\be_3 \\
\be_3 &  & \be_1+\be_2+\be_4 \\
\end{array}\right\}$}

\vspace{9pt}

$D_{23}^4=\left\{ \begin{array}{ccc}
\bzero &  & \be_4 \\
\be_1 &  & \be_2+\be_3 \\
\be_2 &  & \be_2+\be_1+\be_3 \\
\be_3 &  & \be_2+\be_1+\be_4 \\
\end{array}\right\} \leftrightarrow  {\color{Salmon}
D_{22}^4}, P=(12)$

\vspace{9pt}

$D_{24}^4=\left\{ \begin{array}{ccc}
\bzero &  & \be_4 \\
\be_1 &  & \be_2+\be_4 \\
\be_2 &  & \be_2+\be_1+\be_3 \\
\be_3 &  & \be_2+\be_1+\be_4 \\
\end{array}\right\} \leftrightarrow {\color{Salmon}
D_{22}^4}, P=(12)(34)$

\vspace{9pt}

$D_{25}^4=\left\{ \begin{array}{ccc}
\bzero &  & \be_4 \\
\be_1 &  & \be_1+\be_2 \\
\be_2 &  & \be_1+\be_2+\be_3 \\
\be_3 &  & \be_1+\be_3+\be_4 \\
\end{array}\right\} \leftrightarrow {\color{Salmon}
D_{22}^4}, P=(23)$

\vspace{9pt}

$D_{26}^4=\left\{ \begin{array}{ccc}
\bzero &  & \be_4 \\
\be_1 &  & \be_1+\be_3 \\
\be_2 &  & \be_1+\be_3+\be_2 \\
\be_3 &  & \be_1+\be_3+\be_4 \\
\end{array}\right\} \leftrightarrow {\color{Periwinkle}
D_{21}^4}, P=(23)$

\vspace{9pt}

$D_{27}^4=\left\{ \begin{array}{ccc}
\bzero &  & \be_4 \\
\be_1 &  & \be_1+\be_4 \\
\be_2 &  & \be_1+\be_3+\be_2 \\
\be_3 &  & \be_1+\be_3+\be_4 \\
\end{array}\right\} \leftrightarrow {\color{Salmon}
D_{22}^4}, P=(243)$

\vspace{9pt}

$D_{28}^4=\left\{ \begin{array}{ccc}
\bzero &  & \be_4 \\
\be_1 &  & \be_3+\be_4 \\
\be_2 &  & \be_3+\be_2+\be_1 \\
\be_3 &  & \be_3+\be_4+\be_1 \\
\end{array}\right\} \leftrightarrow {\color{Salmon}
D_{22}^4}, P=(1243)$

\vspace{9pt}

$D_{29}^4=\left\{ \begin{array}{ccc}
\bzero &  & \be_4 \\
\be_1 &  & \be_3+\be_2 \\
\be_2 &  & \be_3+\be_1+\be_2 \\
\be_3 &  & \be_3+\be_1+\be_4 \\
\end{array}\right\} \leftrightarrow {\color{Salmon}
D_{22}^4}, P=(123)$

\vspace{9pt}

$D_{30}^4=\left\{ \begin{array}{ccc}
\bzero &  & \be_4 \\
\be_1 &  & \be_2+\be_1 \\
\be_2 &  & \be_2+\be_3+\be_1 \\
\be_3 &  & \be_2+\be_3+\be_4 \\
\end{array}\right\} \leftrightarrow {\color{Salmon}
D_{22}^4}, P=(132)$

\vspace{9pt}

$D_{31}^4=\left\{ \begin{array}{ccc}
\bzero &  & \be_4 \\
\be_1 &  & \be_3+\be_1 \\
\be_2 &  & \be_3+\be_2+\be_1 \\
\be_3 &  & \be_3+\be_2+\be_4 \\
\end{array}\right\} \leftrightarrow {\color{Salmon}
D_{22}^4}, P=(13)$

\vspace{9pt}

$D_{32}^4=\left\{ \begin{array}{ccc}
\bzero &  & \be_4 \\
\be_1 &  & \be_3+\be_4 \\
\be_2 &  & \be_3+\be_2+\be_1 \\
\be_3 &  & \be_3+\be_2+\be_4 \\
\end{array}\right\} \leftrightarrow {\color{Salmon}
D_{22}^4}, P=(143)$

\vspace{9pt}

$D_{33}^4=\left\{ \begin{array}{ccc}
\bzero &  & \be_4 \\
\be_1 &  & \be_2+\be_3 \\
\be_2 &  & \be_2+\be_3+\be_1 \\
\be_3 &  & \be_2+\be_3+\be_4 \\
\end{array}\right\} \leftrightarrow {\color{Periwinkle}
D_{21}^4}, P=(13)$

\vspace{9pt}

$D_{34}^4=\left\{ \begin{array}{ccc}
\bzero &  & \be_4 \\
\be_1 &  & \be_2+\be_4 \\
\be_2 &  & \be_2+\be_3+\be_1 \\
\be_3 &  & \be_2+\be_3+\be_4 \\
\end{array}\right\} \leftrightarrow {\color{Salmon}
D_{22}^4}, P=(1432)$

\vspace{9pt}

$D_{35}^4=\left\{ \begin{array}{ccc}
\bzero &  & \be_4 \\
\be_1 &  & \be_1+\be_2 \\
\be_2 &  & \be_1+\be_4+\be_2 \\
\be_3 &  & \be_1+\be_4+\be_3 \\
\end{array}\right\} \leftrightarrow {\color{Salmon}
D_{22}^4}, P=(234)$

\vspace{9pt}

$D_{36}^4=\left\{ \begin{array}{ccc}
\bzero &  & \be_4 \\
\be_1 &  & \be_1+\be_3 \\
\be_2 &  & \be_1+\be_4+\be_2 \\
\be_3 &  & \be_1+\be_4+\be_3 \\
\end{array}\right\} \leftrightarrow {\color{Salmon}
D_{22}^4}, P=(24)$

\vspace{9pt}

$D_{37}^4=\left\{ \begin{array}{ccc}
\bzero &  & \be_4 \\
\be_1 &  & \be_1+\be_4 \\
\be_2 &  & \be_1+\be_4+\be_2 \\
\be_3 &  & \be_1+\be_4+\be_3 \\
\end{array}\right\} \leftrightarrow {\color{Periwinkle}
D_{21}^4}, P=(24)$

\vspace{9pt}

$D_{38}^4=\left\{ \begin{array}{ccc}
\bzero &  & \be_4 \\
\be_1 &  & \be_4+\be_3 \\
\be_2 &  & \be_4+\be_1+\be_2 \\
\be_3 &  & \be_4+\be_1+\be_3 \\
\end{array}\right\} \leftrightarrow {\color{Salmon}
D_{22}^4}, P=(124)$

\vspace{9pt}

$D_{39}^4=\left\{ \begin{array}{ccc}
\bzero &  & \be_4 \\
\be_1 &  & \be_4+\be_2 \\
\be_2 &  & \be_4+\be_1+\be_2 \\
\be_3 &  & \be_4+\be_1+\be_3 \\
\end{array}\right\} \leftrightarrow {\color{Salmon}
D_{22}^4}, P=(1234)$

\vspace{9pt}

$D_{40}^4=\left\{ \begin{array}{ccc}
\bzero &  & \be_4 \\
\be_1 &  & \be_2+\be_1 \\
\be_2 &  & \be_2+\be_4+\be_1 \\
\be_3 &  & \be_2+\be_4+\be_3 \\
\end{array}\right\} \leftrightarrow {\color{Salmon}
D_{22}^4}, P=(1342)$

\vspace{9pt}

$D_{41}^4=\left\{ \begin{array}{ccc}
\bzero &  & \be_4 \\
\be_1 &  & \be_4+\be_1 \\
\be_2 &  & \be_4+\be_2+\be_1 \\
\be_3 &  & \be_4+\be_2+\be_3 \\
\end{array}\right\} \leftrightarrow {\color{Salmon}
D_{22}^4}, P=(134)$

\vspace{9pt}

$D_{42}^4=\left\{ \begin{array}{ccc}
\bzero &  & \be_4 \\
\be_1 &  & \be_4+\be_3 \\
\be_2 &  & \be_4+\be_2+\be_1 \\
\be_3 &  & \be_4+\be_2+\be_3 \\
\end{array}\right\} \leftrightarrow {\color{Salmon}
D_{22}^4}, P=(14)$

\vspace{9pt}

$D_{43}^4=\left\{ \begin{array}{ccc}
\bzero &  & \be_4 \\
\be_1 &  & \be_2+\be_3 \\
\be_2 &  & \be_2+\be_4+\be_1 \\
\be_3 &  & \be_2+\be_4+\be_3 \\
\end{array}\right\} \leftrightarrow {\color{Salmon}
D_{22}^4}, P=(142)$

\vspace{9pt}

$D_{44}^4=\left\{ \begin{array}{ccc}
\bzero &  & \be_4 \\
\be_1 &  & \be_2+\be_4 \\
\be_2 &  & \be_2+\be_4+\be_1 \\
\be_3 &  & \be_2+\be_4+\be_3 \\
\end{array}\right\} \leftrightarrow {\color{Periwinkle}
D_{21}^4}, P=(14)$

\vspace{9pt}

$D_{45}^4=\left\{ \begin{array}{ccc}
\bzero &  & \be_4 \\
\be_1 &  & \be_3+\be_1 \\
\be_2 &  & \be_3+\be_4+\be_1 \\
\be_3 &  & \be_3+\be_4+\be_2 \\
\end{array}\right\} \leftrightarrow {\color{Salmon}
D_{22}^4}, P=(13)(24)$

\vspace{9pt}

$D_{46}^4=\left\{ \begin{array}{ccc}
\bzero &  & \be_4 \\
\be_1 &  & \be_4+\be_1 \\
\be_2 &  & \be_4+\be_3+\be_1 \\
\be_3 &  & \be_2+\be_3+\be_4 \\
\end{array}\right\} \leftrightarrow {\color{Salmon}
D_{22}^4}, P=(1324)$

\vspace{9pt}

$D_{47}^4=\left\{ \begin{array}{ccc}
\bzero &  & \be_4 \\
\be_1 &  & \be_3+\be_4 \\
\be_2 &  & \be_3+\be_4+\be_1 \\
\be_3 &  & \be_3+\be_4+\be_2 \\
\end{array}\right\} \leftrightarrow {\color{Periwinkle}
D_{21}^4}, P=(13)(24)$

\vspace{9pt}

$D_{48}^4=\left\{ \begin{array}{ccc}
\bzero &  & \be_4 \\
\be_1 &  & \be_3+\be_2 \\
\be_2 &  & \be_3+\be_4+\be_1 \\
\be_3 &  & \be_3+\be_4+\be_2 \\
\end{array}\right\} \leftrightarrow {\color{Salmon}
D_{22}^4}, P=(1423)$

\vspace{9pt}

$D_{49}^4=\left\{ \begin{array}{ccc}
\bzero &  & \be_4 \\
\be_1 &  & \be_4+\be_2 \\
\be_2 &  & \be_4+\be_3+\be_1 \\
\be_3 &  & \be_4+\be_3+\be_2 \\
\end{array}\right\} \leftrightarrow {\color{Salmon}
D_{22}^4}, P=(14)(23)$

\vspace{9pt}

$D_{50}^4=\left\{ \begin{array}{ccc}
\bzero &  & \be_4 \\
\be_1 &  & \be_1+\be_4 \\
\be_2 &  & \be_1+\be_2+\be_3 \\
\be_3 &  & \be_1+\be_2+\be_4 \\
\end{array}\right\} \leftrightarrow {\color{Salmon}
D_{22}^4}, P=(34)$

\vspace{6pt}
Let $D=\{\bzero, \be_1,\be_2,\be_3,\be_4,\bx,\by,\bz\}$, where $\omega_H(\bx)=\omega_H(\by)=\omega_H(\bz)=3$. Then, all the possibilities for $D$ are \\

{\color{OrangeRed}
${\color{OrangeRed} D_{51}^4} =\left\{ \begin{array}{ccc}
\bzero &  & \be_4 \\
\be_1 &  & \be_1+\be_2+\be_3 \\
\be_2 &  & \be_1+\be_2+\be_4 \\
\be_3 &  & \be_1+\be_3+\be_4 \\
\end{array}\right\}  $}

\vspace{9pt}

$D_{52}^4=\left\{ \begin{array}{ccc}
\bzero &  & \be_4 \\
\be_1 &  & \be_2+\be_1+\be_3 \\
\be_2 &  & \be_2+\be_1+\be_4 \\
\be_3 &  & \be_2+\be_3+\be_4 \\
\end{array}\right\} \leftrightarrow {\color{OrangeRed} D_{51}^4} , P=(12)$

\vspace{9pt}

$D_{53}^4=\left\{ \begin{array}{ccc}
\bzero &  & \be_4 \\
\be_1 &  & \be_4+\be_2+\be_1 \\
\be_2 &  & \be_4+\be_3+\be_1 \\
\be_3 &  & \be_4+\be_3+\be_2 \\
\end{array}\right\} \leftrightarrow {\color{OrangeRed} D_{51}^4} , P=(14)$

\vspace{9pt}

$D_{54}^4=\left\{ \begin{array}{ccc}
\bzero &  & \be_4 \\
\be_1 &  & \be_3+\be_2+\be_1 \\
\be_2 &  & \be_3+\be_1+\be_4 \\
\be_3 &  & \be_3+\be_2+\be_4 \\
\end{array}\right\} \leftrightarrow {\color{OrangeRed} D_{51}^4} , P=(13)$

\vspace{9pt}

$\Ran (D)=5$.

\vspace{6pt}
Let $D=\{\bzero, \be_1,\be_2,\be_3,\be_4,\be_5,\bx,\by\}$, where $\omega_H(\bx)=\omega_H(\by)=2$ and $\omega_H(\bx+\by)=2$. Then, all the possibilities for $D$ are \\

{\color{Magenta}
${\color{Magenta} D_{1}^5}=\left\{ \begin{array}{ccc}
0 &  & \be_4 \\
\be_1 &  & \be_5\\
\be_2 &  & \be_1+\be_2 \\
\be_3 &  & \be_1+\be_3 \\
\end{array}\right\}  $}

\vspace{9pt}

$D_{2}^5=\left\{ \begin{array}{ccc}
0 &  & \be_4 \\
\be_1 &  & \be_5\\
\be_2 &  & \be_1+\be_2 \\
\be_3 &  & \be_1+\be_4 \\
\end{array}\right\} \leftrightarrow {\color{Magenta} D_{1}^5}, P=(34)$

\vspace{9pt}

$D_{3}^5=\left\{ \begin{array}{ccc}
0 &  & \be_4 \\
\be_1 &  & \be_5\\
\be_2 &  & \be_1+\be_2 \\
\be_3 &  & \be_1+\be_5 \\
\end{array}\right\} \leftrightarrow {\color{Magenta} D_{1}^5}, P=(35)$

\vspace{9pt}

$D_{4}^5=\left\{ \begin{array}{ccc}
0 &  & \be_4 \\
\be_1 &  & \be_5\\
\be_2 &  & \be_2+\be_1 \\
\be_3 &  & \be_2+\be_3 \\
\end{array}\right\} \leftrightarrow {\color{Magenta} D_{1}^5}, P=(12)$

\vspace{9pt}

$D_{5}^5=\left\{ \begin{array}{ccc}
0 &  & \be_4 \\
\be_1 &  & \be_5\\
\be_2 &  & \be_2+\be_1 \\
\be_3 &  & \be_2+\be_4 \\
\end{array}\right\} \leftrightarrow {\color{Magenta} D_{1}^5}, P=(12)(34)$

\vspace{9pt}

$D_{6}^5=\left\{ \begin{array}{ccc}
0 &  & \be_4 \\
\be_1 &  & \be_5\\
\be_2 &  & \be_2+\be_1 \\
\be_3 &  & \be_2+\be_5 \\
\end{array}\right\} \leftrightarrow {\color{Magenta} D_{1}^5}, P=(12)(35)$

\vspace{9pt}

$D_{7}^5=\left\{ \begin{array}{ccc}
0 &  & \be_4 \\
\be_1 &  & \be_5\\
\be_2 &  & \be_1+\be_3 \\
\be_3 &  & \be_1+\be_4 \\
\end{array}\right\} \leftrightarrow {\color{Magenta} D_{1}^5}, P=(24)$

\vspace{9pt}

$D_{8}^5=\left\{ \begin{array}{ccc}
0 &  & \be_4 \\
\be_1 &  & \be_5\\
\be_2 &  & \be_1+\be_3 \\
\be_3 &  & \be_1+\be_5 \\
\end{array}\right\} \leftrightarrow {\color{Magenta} D_{1}^5}, P=(25)$

\vspace{9pt}

$D_{9}^5=\left\{ \begin{array}{ccc}
0 &  & \be_4 \\
\be_1 &  & \be_5\\
\be_2 &  & \be_3+\be_1 \\
\be_3 &  & \be_3+\be_2 \\
\end{array}\right\} \leftrightarrow {\color{Magenta} D_{1}^5}, P=(13)$

\vspace{9pt}

$D_{10}^5=\left\{ \begin{array}{ccc}
0 &  & \be_4 \\
\be_1 &  & \be_5\\
\be_2 &  & \be_3+\be_1 \\
\be_3 &  & \be_3+\be_4 \\
\end{array}\right\} \leftrightarrow {\color{Magenta} D_{1}^5}, P=(13)(24)$

\vspace{9pt}

$D_{11}^5=\left\{ \begin{array}{ccc}
0 &  & \be_4 \\
\be_1 &  & \be_5\\
\be_2 &  & \be_3+\be_1 \\
\be_3 &  & \be_3+\be_5 \\
\end{array}\right\} \leftrightarrow {\color{Magenta} D_{1}^5}, P=(13)(25)$

\vspace{9pt}

$D_{12}^5=\left\{ \begin{array}{ccc}
0 &  & \be_4 \\
\be_1 &  & \be_5\\
\be_2 &  & \be_1+\be_4 \\
\be_3 &  & \be_1+\be_5 \\
\end{array}\right\} \leftrightarrow {\color{Magenta} D_{1}^5}, P=(24)(35)$

\vspace{9pt}

$D_{13}^5=\left\{ \begin{array}{ccc}
0 &  & \be_4 \\
\be_1 &  & \be_5\\
\be_2 &  & \be_4+\be_1 \\
\be_3 &  & \be_4+\be_2 \\
\end{array}\right\} \leftrightarrow {\color{Magenta} D_{1}^5}, P=(134)$

\vspace{9pt}

$D_{14}^5=\left\{ \begin{array}{ccc}
0 &  & \be_4 \\
\be_1 &  & \be_5\\
\be_2 &  & \be_4+\be_1 \\
\be_3 &  & \be_4+\be_5 \\
\end{array}\right\} \leftrightarrow {\color{Magenta} D_{1}^5}, P=(124)(35)$

\vspace{9pt}

$D_{15}^5=\left\{ \begin{array}{ccc}
0 &  & \be_4 \\
\be_1 &  & \be_5\\
\be_2 &  & \be_5+\be_1 \\
\be_3 &  & \be_5+\be_2 \\
\end{array}\right\} \leftrightarrow {\color{Magenta} D_{1}^5}, P=(135)$

\vspace{9pt}

$D_{16}^5=\left\{ \begin{array}{ccc}
0 &  & \be_4 \\
\be_1 &  & \be_5\\
\be_2 &  & \be_5+\be_1 \\
\be_3 &  & \be_5+\be_3 \\
\end{array}\right\} \leftrightarrow {\color{Magenta} D_{1}^5}, P=(125)$

\vspace{9pt}

$D_{17}^5=\left\{ \begin{array}{ccc}
0 &  & \be_4 \\
\be_1 &  & \be_5\\
\be_2 &  & \be_5+\be_1 \\
\be_3 &  & \be_5+\be_4 \\
\end{array}\right\} \leftrightarrow {\color{Magenta} D_{1}^5}, P=(125)(34)$

\vspace{9pt}

$D_{18}^5=\left\{ \begin{array}{ccc}
0 &  & \be_4 \\
\be_1 &  & \be_5\\
\be_2 &  & \be_2+\be_3 \\
\be_3 &  & \be_2+\be_4 \\
\end{array}\right\} \leftrightarrow {\color{Magenta} D_{1}^5}, P=(142)$

\vspace{9pt}

$D_{19}^5=\left\{ \begin{array}{ccc}
0 &  & \be_4 \\
\be_1 &  & \be_5\\
\be_2 &  & \be_2+\be_3 \\
\be_3 &  & \be_2+\be_5 \\
\end{array}\right\} \leftrightarrow {\color{Magenta} D_{1}^5}, P=(152)$

\vspace{9pt}

$D_{20}^5=\left\{ \begin{array}{ccc}
0 &  & \be_4 \\
\be_1 &  & \be_5\\
\be_2 &  & \be_3+\be_2 \\
\be_3 &  & \be_3+\be_4 \\
\end{array}\right\} \leftrightarrow {\color{Magenta} D_{1}^5}, P=(143)$

\vspace{9pt}

$D_{21}^5=\left\{ \begin{array}{ccc}
0 &  & \be_4 \\
\be_1 &  & \be_5\\
\be_2 &  & \be_3+\be_2 \\
\be_3 &  & \be_3+\be_5 \\
\end{array}\right\} \leftrightarrow {\color{Magenta} D_{1}^5}, P=(153)$

\vspace{9pt}

$D_{22}^5=\left\{ \begin{array}{ccc}
0 &  & \be_4 \\
\be_1 &  & \be_5\\
\be_2 &  & \be_2+\be_4 \\
\be_3 &  & \be_2+\be_5 \\
\end{array}\right\} \leftrightarrow {\color{Magenta} D_{1}^5}, P=(142)(35)$

\vspace{9pt}

$D_{23}^5=\left\{ \begin{array}{ccc}
0 &  & \be_4 \\
\be_1 &  & \be_5\\
\be_2 &  & \be_4+\be_2 \\
\be_3 &  & \be_4+\be_3 \\
\end{array}\right\} \leftrightarrow {\color{Magenta} D_{1}^5}, P=(14)$

\vspace{9pt}

$D_{24}^5=\left\{ \begin{array}{ccc}
0 &  & \be_4 \\
\be_1 &  & \be_5\\
\be_2 &  & \be_4+\be_2 \\
\be_3 &  & \be_4+\be_5 \\
\end{array}\right\} \leftrightarrow {\color{Magenta} D_{1}^5}, P=(14)(35)$

\vspace{9pt}

$D_{25}^5=\left\{ \begin{array}{ccc}
0 &  & \be_4 \\
\be_1 &  & \be_5\\
\be_2 &  & \be_5+\be_2 \\
\be_3 &  & \be_5+\be_3 \\
\end{array}\right\} \leftrightarrow {\color{Magenta} D_{1}^5}, P=(15)$

\vspace{9pt}

$D_{26}^5=\left\{ \begin{array}{ccc}
0 &  & \be_4 \\
\be_1 &  & \be_5\\
\be_2 &  & \be_5+\be_2 \\
\be_3 &  & \be_5+\be_4 \\
\end{array}\right\} \leftrightarrow {\color{Magenta} D_{1}^5}, P=(15)(34)$

\vspace{9pt}

$D_{27}^5=\left\{ \begin{array}{ccc}
0 &  & \be_4 \\
\be_1 &  & \be_5\\
\be_2 &  & \be_3+\be_4 \\
\be_3 &  & \be_3+\be_5 \\
\end{array}\right\} \leftrightarrow {\color{Magenta} D_{1}^5}, P=(153)(24)$
\vspace{9pt}

$D_{28}^5=\left\{ \begin{array}{ccc}
0 &  & \be_4 \\
\be_1 &  & \be_5\\
\be_2 &  & \be_4+\be_3 \\
\be_3 &  & \be_4+\be_5 \\
\end{array}\right\} \leftrightarrow {\color{Magenta} D_{1}^5}, P=(14)(25)$

\vspace{9pt}

$D_{29}^5=\left\{ \begin{array}{ccc}
0 &  & \be_4 \\
\be_1 &  & \be_5\\
\be_2 &  & \be_5+\be_3 \\
\be_3 &  & \be_5+\be_4 \\
\end{array}\right\} \leftrightarrow {\color{Magenta} D_{1}^5}, P=(15)(24)$

$D_{30}^5=\left\{ \begin{array}{ccc}
0 &  & \be_4 \\
\be_1 &  & \be_5\\
\be_2 &  & \be_4+\be_1 \\
\be_3 &  & \be_4+\be_3 \\
\end{array}\right\} \leftrightarrow {\color{Magenta} D_{1}^5}, P=(124)$

$D_{31}^5=\left\{ \begin{array}{ccc}
0 &  & \be_4 \\
\be_1 &  & \be_5\\
\be_2 &  & \be_1+\be_4 \\
\be_3 &  & \be_1+\be_5 \\
\end{array}\right\} \leftrightarrow {\color{Magenta} D_{1}^5}, P=(24)(35)$

\vspace{6pt}
Let $D=\{\bzero, \be_1,\be_2,\be_3,\be_4,\be_5,\bx,\by\}$, where $\omega_H(\bx)=3$, $\omega_H(\by)=2$ and $\omega_H(\bx+\by)=1$ or $\omega_H(\bx+\by)=5$. Then, all the possibilities for $D$ are \\

{\color{NavyBlue}
${\color{NavyBlue} D_{32}^5}=\left\{ \begin{array}{cc}
0 &   \be_4 \\
\be_1 &   \be_5\\
\be_2 &  \be_1+\be_2 \\
\be_3 &   \be_1+\be_2+\be_3 \\
\end{array}\right\} $}

\vspace{9pt}

{\color{RedOrange}
${\color{RedOrange} D_{33}^5}=\left\{ \begin{array}{cc}
0 &   \be_4 \\
\be_1   & \be_5\\
\be_2   & \be_4+\be_5 \\
\be_3   & \be_1+\be_2+\be_3 \\
\end{array}\right\}  $}

\vspace{9pt}

$D_{34}^5=\left\{ \begin{array}{cc}
0 &   \be_4 \\
\be_1   & \be_5\\
\be_2   & \be_1+\be_2 \\
\be_3   & \be_1+\be_2+\be_4 \\
\end{array}\right\} \leftrightarrow  {\color{NavyBlue} D_{32}^5}, P=(34)$

\vspace{9pt}

$D_{35}^5=\left\{ \begin{array}{cc}
0 &   \be_4 \\
\be_1  & \be_5\\
\be_2 & \be_1+\be_4 \\
\be_3 & \be_1+\be_2+\be_4 \\
\end{array}\right\} \leftrightarrow  {\color{NavyBlue} D_{32}^5},P=(234)$

\vspace{9pt}

$D_{36}^5=\left\{ \begin{array}{cc}
0 &  \be_4 \\
\be_1  & \be_5\\
\be_2  & \be_2+\be_4 \\
\be_3 & \be_2+\be_4+\be_1 \\
\end{array}\right\} \leftrightarrow  {\color{NavyBlue} D_{32}^5}, P=(134)$

\vspace{9pt}

$D_{37}^5=\left\{ \begin{array}{cc}
0 &   \be_4 \\
\be_1   & \be_5\\
\be_2   & \be_3+\be_5 \\
\be_3   & \be_1+\be_2+\be_4 \\
\end{array}\right\} \leftrightarrow  {\color{RedOrange} D_{33}^5}, P=(34)$

\vspace{9pt}

$D_{38}^5=\left\{ \begin{array}{cc}
0 &   \be_4 \\
\be_1   & \be_5\\
\be_2   & \be_1+\be_2 \\
\be_3   & \be_1+\be_2+\be_5 \\
\end{array}\right\} \leftrightarrow  {\color{NavyBlue} D_{32}^5}, P=(35)$
\vspace{9pt}

$D_{39}^5=\left\{ \begin{array}{cc}
0 &   \be_4 \\
\be_1  & \be_5\\
\be_2  & \be_1+\be_5 \\
\be_3  & \be_1+\be_5+\be_2 \\
\end{array}\right\} \leftrightarrow  {\color{NavyBlue} D_{32}^5}, P=(235)$

\vspace{9pt}

$D_{40}^5=\left\{ \begin{array}{cc}
0 &   \be_4 \\
\be_1   & \be_5\\
\be_2   & \be_2+\be_5 \\
\be_3   & \be_2+\be_5+\be_1 \\
\end{array}\right\} \leftrightarrow  {\color{NavyBlue} D_{32}^5}, P=(135)$

\vspace{9pt}

$D_{41}^5=\left\{ \begin{array}{cc}
0 &   \be_4 \\
\be_1 &   \be_5\\
\be_2 &   \be_3+\be_4 \\
\be_3 &   \be_1+\be_2+\be_5 \\
\end{array}\right\} \leftrightarrow  {\color{RedOrange} D_{33}^5}, P=(35)$

\vspace{9pt}

$D_{42}^5=\left\{ \begin{array}{cc}
0 &   \be_4 \\
\be_1   & \be_5\\
\be_2   & \be_1+\be_3 \\
\be_3   & \be_1+\be_3+\be_4 \\
\end{array}\right\} \leftrightarrow  {\color{NavyBlue} D_{32}^5}, P=(243)$

\vspace{9pt}

$D_{43}^5=\left\{ \begin{array}{cc}
0 &   \be_4 \\
\be_1   & \be_5\\
\be_2   & \be_1+\be_4 \\
\be_3   & \be_1+\be_4+\be_3 \\
\end{array}\right\} \leftrightarrow  {\color{NavyBlue} D_{32}^5}, P=(234)$

\vspace{9pt}

$D_{44}^5=\left\{ \begin{array}{cc}
0 &   \be_4 \\
\be_1   & \be_5\\
\be_2   & \be_2+\be_5 \\
\be_3   & \be_1+\be_3+\be_4 \\
\end{array}\right\} \leftrightarrow  {\color{RedOrange} D_{33}^5}, P=(243)$

\vspace{9pt}

$D_{45}^5=\left\{ \begin{array}{cc}
0 &   \be_4 \\
\be_1   & \be_5\\
\be_2   & \be_3+\be_4 \\
\be_3   & \be_3+\be_4+\be_1 \\
\end{array}\right\} \leftrightarrow {\color{NavyBlue} D_{32}^5}, P=(13)(24)$

\vspace{9pt}

$D_{46}^5=\left\{ \begin{array}{cc}
0 &   \be_4 \\
\be_1 &   \be_5\\
\be_2 &   \be_1+\be_4 \\
\be_3 &   \be_1+\be_4+\be_5 \\
\end{array}\right\} \leftrightarrow {\color{NavyBlue} D_{32}^5}, P=(24)(35)$

\vspace{9pt}

$D_{47}^5=\left\{ \begin{array}{cc}
0 &   \be_4 \\
\be_1 &   \be_5\\
\be_2 &  \be_1+\be_5 \\
\be_3 &   \be_1+\be_5+\be_4 \\
\end{array}\right\} \leftrightarrow {\color{NavyBlue} D_{32}^5}, P=(25)(34)$

\vspace{9pt}

$D_{48}^5=\left\{ \begin{array}{cc}
0 &   \be_4 \\
\be_1   & \be_5\\
\be_2   & \be_2+\be_3 \\
\be_3   & \be_1+\be_4+\be_5 \\
\end{array}\right\} \leftrightarrow {\color{RedOrange} D_{33}^5}, P=(24)(35)$

\vspace{9pt}

$D_{49}^5=\left\{ \begin{array}{cc}
0 &   \be_4 \\
\be_1   & \be_5\\
\be_2   & \be_4+\be_5 \\
\be_3   & \be_4+\be_5+\be_1 \\
\end{array}\right\} \leftrightarrow {\color{NavyBlue} D_{32}^5}, P=(134)(25)$

\vspace{9pt}

$D_{50}^5=\left\{ \begin{array}{cc}
0 &   \be_4 \\
\be_1   & \be_5\\
\be_2   & \be_1+\be_3 \\
\be_3   & \be_1+\be_3+\be_5 \\
\end{array}\right\} \leftrightarrow  {\color{NavyBlue} D_{32}^5}, P=(253)$

\vspace{9pt}

$D_{51}^5=\left\{ \begin{array}{cc}
0 &   \be_4 \\
\be_1   & \be_5\\
\be_2   & \be_1+\be_5 \\
\be_3   & \be_1+\be_5+\be_3 \\
\end{array}\right\} \leftrightarrow  {\color{NavyBlue} D_{32}^5}, P=(25)$

\vspace{9pt}

$D_{52}^5=\left\{ \begin{array}{cc}
0 &   \be_4 \\
\be_1   & \be_5\\
\be_2   & \be_2+\be_4 \\
\be_3   & \be_1+\be_3+\be_5 \\
\end{array}\right\} \leftrightarrow  {\color{RedOrange} D_{33}^5}, P=(25)$

\vspace{9pt}

$D_{53}^5=\left\{ \begin{array}{cc}
0 &   \be_4 \\
\be_1   & \be_5\\
\be_2   & \be_3+\be_5 \\
\be_3   & \be_3+\be_5+\be_1 \\
\end{array}\right\} \leftrightarrow {\color{NavyBlue} D_{32}^5}, P=(13)(25)$

\vspace{9pt}

$D_{54}^5=\left\{ \begin{array}{cc}
0 &   \be_4 \\
\be_1   & \be_5\\
\be_2   & \be_1+\be_5 \\
\be_3   & \be_2+\be_3+\be_4 \\
\end{array}\right\} \leftrightarrow {\color{RedOrange} D_{33}^5}, P=(14)
$

\vspace{9pt}

$D_{55}^5=\left\{ \begin{array}{cc}
0 & \be_4 \\
\be_1   & \be_5\\
\be_2   & \be_2+\be_3 \\
\be_3   & \be_2+\be_3+\be_4 \\
\end{array}\right\} \leftrightarrow {\color{NavyBlue} D_{32}^5}, P=(143)$

\vspace{9pt}

$D_{56}^5=\left\{ \begin{array}{cc}
0 &   \be_4 \\
\be_1   & \be_5\\
\be_2   & \be_2+\be_4 \\
\be_3   & \be_2+\be_4+\be_3 \\
\end{array}\right\} \leftrightarrow {\color{NavyBlue} D_{32}^5}, P=(14)$

\vspace{9pt}

$D_{57}^5=\left\{ \begin{array}{cc}
0  & \be_4 \\
\be_1  & \be_5\\
\be_2   & \be_3+\be_4 \\
\be_3   & \be_3+\be_4+\be_2 \\
\end{array}\right\} \leftrightarrow {\color{NavyBlue} D_{32}^5}, P=(1423)$
\vspace{9pt}

$D_{58}^5=\left\{ \begin{array}{cc}
0 &   \be_4 \\
\be_1   & \be_5\\
\be_2   & \be_1+\be_2 \\
\be_3   & \be_3+\be_4+\be_5 \\
\end{array}\right\} \leftrightarrow {\color{RedOrange} D_{33}^5}, P=(14)(25)$

\vspace{9pt}

$D_{59}^5=\left\{ \begin{array}{cc}
0 &   \be_4 \\
\be_1   & \be_5\\
\be_2   & \be_3+\be_4 \\
\be_3   & \be_3+\be_4+\be_5 \\
\end{array}\right\} \leftrightarrow {\color{NavyBlue} D_{32}^5}, P=(143)(25)$

\vspace{9pt}

$D_{60}^5=\left\{ \begin{array}{cc}
0 &   \be_4 \\
\be_1   & \be_5\\
\be_2   & \be_3+\be_5 \\
\be_3   & \be_3+\be_5+\be_4 \\
\end{array}\right\} \leftrightarrow {\color{NavyBlue} D_{32}^5}, P=(143)(25)$

\vspace{9pt}

$D_{61}^5=\left\{ \begin{array}{cc}
0 &   \be_4 \\
\be_1   & \be_5\\
\be_2   & \be_4+\be_5 \\
\be_3   & \be_4+\be_5+\be_3 \\
\end{array}\right\} \leftrightarrow {\color{NavyBlue} D_{32}^5}, P=(14)(25)$

\vspace{9pt}

$D_{62}^5=\left\{ \begin{array}{cc}
0 &   \be_4 \\
\be_1   & \be_5\\
\be_2   & \be_1+\be_3 \\
\be_3   & \be_2+\be_4+\be_5 \\
\end{array}\right\} \leftrightarrow {\color{RedOrange} D_{33}^5}, P=(14)(35)$

\vspace{9pt}

$D_{63}^5=\left\{ \begin{array}{cc}
0 &   \be_4 \\
\be_1   & \be_5\\
\be_2   & \be_2+\be_4 \\
\be_3   & \be_2+\be_4+\be_5 \\
\end{array}\right\} \leftrightarrow {\color{NavyBlue} D_{32}^5}, P=(14)(35)$
\vspace{9pt}

$D_{64}^5=\left\{ \begin{array}{cc}
0 &   \be_4 \\
\be_1   & \be_5\\
\be_2   & \be_2+\be_5 \\
\be_3   & \be_2+\be_5+\be_4 \\
\end{array}\right\} \leftrightarrow {\color{NavyBlue} D_{32}^5}, P=(15)(34)$

\vspace{9pt}

$D_{65}^5=\left\{ \begin{array}{cc}
0 &   \be_4 \\
\be_1   & \be_5\\
\be_2   & \be_4+\be_5 \\
\be_3   & \be_4+\be_5+\be_2 \\
\end{array}\right\} \leftrightarrow {\color{NavyBlue} D_{32}^5}, P=(14)(235)$

\vspace{9pt}

$D_{66}^5=\left\{ \begin{array}{cc}
0 &   \be_4 \\
\be_1   & \be_5\\
\be_2   & \be_1+\be_4 \\
\be_3   & \be_2+\be_3+\be_5 \\
\end{array}\right\} \leftrightarrow {\color{RedOrange} D_{33}^5}, P=(15)$

\vspace{9pt}

$D_{67}^5=\left\{ \begin{array}{cc}
0 &   \be_4 \\
\be_1   & \be_5\\
\be_2   & \be_2+\be_3 \\
\be_3   & \be_2+\be_3+\be_5 \\
\end{array}\right\} \leftrightarrow {\color{NavyBlue} D_{32}^5}, P=(153)$

\vspace{9pt}

$D_{68}^5=\left\{ \begin{array}{cc}
0 &   \be_4 \\
\be_1   & \be_5\\
\be_2   & \be_2+\be_5 \\
\be_3   & \be_2+\be_5+\be_3 \\
\end{array}\right\} \leftrightarrow {\color{NavyBlue} D_{32}^5}, P=(15)$
\vspace{9pt}

$D_{69}^5=\left\{ \begin{array}{cc}
0 &   \be_4 \\
\be_1   & \be_5\\
\be_2   & \be_3+\be_5 \\
\be_3   & \be_3+\be_5+\be_2 \\
\end{array}\right\} \leftrightarrow {\color{NavyBlue} D_{32}^5}, P=(1523)$

\vspace{9pt}

$D_{70}^5=\left\{ \begin{array}{cc}
0 &   \be_4 \\
\be_1 &   \be_5\\
\be_2 &   \be_1+\be_3 \\
\be_3 &   \be_1+\be_3+\be_2 \\
\end{array}\right\} \leftrightarrow {\color{NavyBlue} D_{32}^5}, P=(23)$

\vspace{9pt}

$D_{71}^5=\left\{ \begin{array}{cc}
0 &   \be_4 \\
\be_1 &   \be_5\\
\be_2 &   \be_2+\be_3 \\
\be_3 &   \be_2+\be_3+\be_1 \\
\end{array}\right\} \leftrightarrow  {\color{NavyBlue} D_{32}^5}, P=(13)$

\vspace{9pt}

Let $D=\{\bzero, \be_1,\be_2,\be_3,\be_4,\be_5,\bx,\by\}$, where $\omega_H(\bx)=\omega_H(\by)=3$ and $\omega_H(\bx+\by)=2$. Then, all the possibilities for $D$ are \\

\vspace{6pt}

{\color{Mahogany}
$D_{72}^5=\left\{ \begin{array}{cc}
0 &   \be_4 \\
\be_1   & \be_5\\
\be_2   & \be_1+\be_2+\be_3 \\
\be_3   & \be_1+\be_2+\be_4 \\
\end{array}\right\}  $}

\vspace{9pt}

$D_{73}^5=\left\{ \begin{array}{cc}
0 &   \be_4 \\
\be_1   & \be_5\\
\be_2   & \be_1+\be_3+\be_2 \\
\be_3   & \be_1+\be_3+\be_5 \\
\end{array}\right\} \leftrightarrow {\color{Mahogany}
D_{72}^5}, P=(23)(45)$
\vspace{9pt}

$D_{74}^5=\left\{ \begin{array}{cc}
0 &   \be_4 \\
\be_1   & \be_5\\
\be_2   & \be_2+\be_3+\be_1 \\
\be_3   & \be_2+\be_3+\be_4 \\
\end{array}\right\} \leftrightarrow {\color{Mahogany}
D_{72}^5}, P=(13)$

\vspace{9pt}

$D_{75}^5=\left\{ \begin{array}{cc}
0 &   \be_4 \\
\be_1   & \be_5\\
\be_2   & \be_2+\be_3+\be_5 \\
\be_3   & \be_2+\be_3+\be_1 \\
\end{array}\right\} \leftrightarrow {\color{Mahogany}
D_{72}^5}, P=(13)(45)$

\vspace{9pt}

$D_{76}^5=\left\{ \begin{array}{cc}
0 &   \be_4 \\
\be_1   & \be_5\\
\be_2   & \be_1+\be_2+\be_4 \\
\be_3   & \be_1+\be_2+\be_5 \\
\end{array}\right\} \leftrightarrow {\color{Mahogany}
D_{72}^5}, P=(35)$

\vspace{9pt}

$D_{77}^5=\left\{ \begin{array}{cc}
0 &   \be_4 \\
\be_1   & \be_5\\
\be_2   & \be_1+\be_4+\be_2 \\
\be_3   & \be_1+\be_4+\be_3 \\
\end{array}\right\} \leftrightarrow {\color{Mahogany}
D_{72}^5}, P=(24)$

\vspace{9pt}

$D_{78}^5=\left\{ \begin{array}{cc}
0 &   \be_4 \\
\be_1   & \be_5\\
\be_2   & \be_1+\be_4+\be_2 \\
\be_3   & \be_1+\be_4+\be_5 \\
\end{array}\right\} \leftrightarrow {\color{Mahogany}
D_{72}^5}, P=(2354)$

\vspace{9pt}

$D_{79}^5=\left\{ \begin{array}{cc}
0 &   \be_4 \\
\be_1   & \be_5\\
\be_2  & \be_2+\be_4+\be_1 \\
\be_3   & \be_2+\be_4+\be_3 \\
\end{array}\right\} \leftrightarrow {\color{Mahogany}
D_{72}^5}, P=(14)$

\vspace{9pt}

$D_{80}^5=\left\{ \begin{array}{cc}
0 &   \be_4 \\
\be_1   & \be_5\\
\be_2   & \be_1+\be_5+\be_2 \\
\be_3   & \be_1+\be_5+\be_4 \\
\end{array}\right\} \leftrightarrow {\color{Mahogany}
D_{72}^5}, P=(235)$

\vspace{9pt}

$D_{81}^5=\left\{ \begin{array}{cc}
0 &   \be_4 \\
\be_1   & \be_5\\
\be_2   & \be_1+\be_5+\be_2 \\
\be_3   & \be_1+\be_5+\be_3 \\
\end{array}\right\} \leftrightarrow {\color{Mahogany}
D_{72}^5}, P=(245)$

\vspace{9pt}

$D_{82}^5=\left\{ \begin{array}{cc}
0 &   \be_4 \\
\be_1   & \be_5\\
\be_2   & \be_2+\be_5+\be_1 \\
\be_3   & \be_2+\be_5+\be_4 \\
\end{array}\right\} \leftrightarrow {\color{Mahogany}
D_{72}^5}, P=(135)$

\vspace{9pt}

$D_{83}^5=\left\{ \begin{array}{cc}
0 &   \be_4 \\
\be_1   & \be_5\\
\be_2   & \be_2+\be_5+\be_1 \\
\be_3   & \be_2+\be_5+\be_3 \\
\end{array}\right\} \leftrightarrow {\color{Mahogany}
D_{72}^5}, P=(145)$

\vspace{9pt}

$D_{84}^5=\left\{ \begin{array}{cc}
0 &   \be_4 \\
\be_1   & \be_5\\
\be_2   & \be_1+\be_4+\be_3 \\
\be_3   & \be_1+\be_4+\be_5 \\
\end{array}\right\} \leftrightarrow {\color{Mahogany}
D_{72}^5}, P=(254)$
\vspace{9pt}

$D_{85}^5=\left\{ \begin{array}{cc}
0 &   \be_4 \\
\be_1   & \be_5\\
\be_2   & \be_1+\be_3+\be_4 \\
\be_3   & \be_1+\be_3+\be_5 \\
\end{array}\right\} \leftrightarrow {\color{Mahogany}
D_{72}^5}, P=(253)$

\vspace{9pt}

$D_{86}^5=\left\{ \begin{array}{cc}
0 &   \be_4 \\
\be_1   & \be_5\\
\be_2   & \be_3+\be_4+\be_1 \\
\be_3   & \be_3+\be_4+\be_2 \\
\end{array}\right\} \leftrightarrow {\color{Mahogany}
D_{72}^5}, P=(13)(24)$
\vspace{9pt}

$D_{87}^5=\left\{ \begin{array}{cc}
0 &   \be_4 \\
\be_1   & \be_5\\
\be_2   & \be_3+\be_4+\be_1 \\
\be_3   & \be_3+\be_4+\be_5 \\
\end{array}\right\} \leftrightarrow {\color{Mahogany}
D_{72}^5}, P=(13)(254)$

\vspace{9pt}

$D_{88}^5=\left\{ \begin{array}{cc}
0 &   \be_4 \\
\be_1   & \be_5\\
\be_2   & \be_1+\be_5+\be_4 \\
\be_3   & \be_1+\be_5+\be_3 \\
\end{array}\right\} \leftrightarrow {\color{Mahogany}
D_{72}^5}, P=(25)$

\vspace{9pt}

$D_{89}^5=\left\{ \begin{array}{cc}
0 &   \be_4 \\
\be_1   & \be_5\\
\be_2   & \be_4+\be_5+\be_1 \\
\be_3   & \be_4+\be_5+\be_3 \\
\end{array}\right\} \leftrightarrow {\color{Mahogany}
D_{72}^5}, P=(14)(25)$

\vspace{9pt}

$D_{90}^5=\left\{ \begin{array}{cc}
0 &   \be_4 \\
\be_1   & \be_5\\
\be_2   & \be_4+\be_5+\be_1 \\
\be_3   & \be_4+\be_5+\be_2 \\
\end{array}\right\} \leftrightarrow {\color{Mahogany}
D_{72}^5}, P=(13524)$

\vspace{9pt}

$D_{91}^5=\left\{ \begin{array}{cc}
0 &   \be_4 \\
\be_1   & \be_5\\
\be_2   & \be_3+\be_5+\be_1 \\
\be_3   & \be_3+\be_5+\be_4 \\
\end{array}\right\} \leftrightarrow {\color{Mahogany}
D_{72}^5}, P=(13)(25)$

\vspace{9pt}

$D_{92}^5=\left\{ \begin{array}{cc}
0 &   \be_4 \\
\be_1   & \be_5\\
\be_2   & \be_3+\be_5+\be_1 \\
\be_3   & \be_3+\be_5+\be_2 \\
\end{array}\right\} \leftrightarrow {\color{Mahogany}
D_{72}^5}, P=(13)(245)$
\vspace{9pt}

$D_{93}^5=\left\{ \begin{array}{cc}
0 &   \be_4 \\
\be_1   & \be_5\\
\be_2   & \be_3+\be_4+\be_2 \\
\be_3   & \be_3+\be_4+\be_5 \\
\end{array}\right\} \leftrightarrow {\color{Mahogany}
D_{72}^5}, P=(153)(24)$

\vspace{9pt}

$D_{94}^5=\left\{ \begin{array}{cc}
0 &   \be_4 \\
\be_1   & \be_5\\
\be_2   & \be_2+\be_4+\be_3 \\
\be_3   & \be_2+\be_4+\be_5 \\
\end{array}\right\} \leftrightarrow {\color{Mahogany}
D_{72}^5}, P=(154)$

\vspace{9pt}

$D_{95}^5=\left\{ \begin{array}{cc}
0 &   \be_4 \\
\be_1   & \be_5\\
\be_2   & \be_2+\be_3+\be_4 \\
\be_3   & \be_2+\be_3+\be_5 \\
\end{array}\right\} \leftrightarrow {\color{Mahogany}
D_{72}^5}, P=(153)$

\vspace{9pt}

$D_{96}^5=\left\{ \begin{array}{cc}
0 &   \be_4 \\
\be_1   & \be_5\\
\be_2   & \be_4+\be_5+\be_3 \\
\be_3   & \be_4+\be_5+\be_2 \\
\end{array}\right\} \leftrightarrow {\color{Mahogany}
D_{72}^5}, P=(1524)$
\vspace{9pt}

$D_{97}^5=\left\{ \begin{array}{cc}
0 &   \be_4 \\
\be_1 &   \be_5\\
\be_2 &   \be_3+\be_5+\be_4 \\
\be_3 &   \be_3+\be_5+\be_2 \\
\end{array}\right\} \leftrightarrow {\color{Mahogany}
D_{72}^5}, P=(1523)$
\vspace{9pt}

$D_{98}^5=\left\{ \begin{array}{cc}
0 &   \be_4 \\
\be_1   & \be_5\\
\be_2   & \be_2+\be_5+\be_4 \\
\be_3   & \be_2+\be_5+\be_3 \\
\end{array}\right\} \leftrightarrow {\color{Mahogany}
D_{72}^5}, P=(15)$

\vspace{6pt}

$D_{99}^5=\left\{ \begin{array}{cc}
0 &   \be_4 \\
\be_1   & \be_5\\
\be_2   & \be_1+\be_2+\be_3 \\
\be_3   & \be_1+\be_2+\be_5 \\
\end{array}\right\} \leftrightarrow {\color{Mahogany}
D_{72}^5}, P=(45)$

\vspace{9pt}

$D_{100}^5=\left\{ \begin{array}{cc}
0 &   \be_4 \\
\be_1   & \be_5\\
\be_2 &   \be_1+\be_3+\be_2 \\
\be_3 &   \be_1+\be_3+\be_4 \\
\end{array}\right\} \leftrightarrow {\color{Mahogany}
D_{72}^5}, P=(23)$

\vspace{9pt}

$D_{101}^5=\left\{ \begin{array}{cc}
0 &   \be_4 \\
\be_1   & \be_5\\
\be_2   & \be_2+\be_4+\be_1 \\
\be_3   & \be_2+\be_4+\be_5\\
\end{array}\right\} \leftrightarrow {\color{Mahogany}
D_{72}^5}, P=(1354)$

\vspace{9pt}

$\Ran (D)=6$.

\vspace{6pt}
Let $D=\{\bzero, \be_1,\be_2,\be_3,\be_4,e_5,e_6,\bx\}$, where $\omega_H(\bx)\not= 4,5$. Then, all the possibilities for $D$ are \\

{\color{ProcessBlue}
${\color{ProcessBlue} {\color{ProcessBlue} D_1^6}}=\left\{ \begin{array}{ccc}
\bzero &  & \be_4 \\
\be_1 &  & \be_5 \\
\be_2 &  & \be_6 \\
\be_3 &  & \be_1+\be_2 \\
\end{array}\right\}  $}

\vspace{9pt}

$D_2^6=\left\{ \begin{array}{ccc}
\bzero &  & \be_4 \\
\be_1 &  & \be_5 \\
\be_2 &  & \be_6 \\
\be_3 &  & \be_1+\be_3 \\
\end{array}\right\} \leftrightarrow {\color{ProcessBlue} D_1^6}, P=(23)$

\vspace{9pt}

$D_3^6=\left\{ \begin{array}{ccc}
0\bzero &  & \be_4 \\
\be_1 &  & \be_5 \\
\be_2 &  & \be_6 \\
\be_3 &  & \be_1+\be_4 \\
\end{array}\right\} \leftrightarrow {\color{ProcessBlue} D_1^6}, P=(24)$

\vspace{9pt}

$D_4^6=\left\{ \begin{array}{ccc}
\bzero &  & \be_4 \\
\be_1 &  & \be_5 \\
\be_2 &  & \be_6 \\
\be_3 &  & \be_1+\be_5 \\
\end{array}\right\} \leftrightarrow {\color{ProcessBlue} D_1^6}, P=(25)$

\vspace{9pt}

$D_5^6=\left\{ \begin{array}{ccc}
\bzero &  & \be_4 \\
\be_1 &  & \be_5 \\
\be_2 &  & \be_6 \\
\be_3 &  & \be_1+\be_6 \\
\end{array}\right\} \leftrightarrow {\color{ProcessBlue} D_1^6}, P=(26)$
\vspace{9pt}

$D_6^6=\left\{ \begin{array}{ccc}
\bzero &  & \be_4 \\
\be_1 &  & \be_5 \\
\be_2 &  & \be_6 \\
\be_3 &  & \be_2+\be_3 \\
\end{array}\right\} \leftrightarrow {\color{ProcessBlue} D_1^6}, P=(13)$

\vspace{9pt}

$D_7^6=\left\{ \begin{array}{ccc}
\bzero &  & \be_4 \\
\be_1 &  & \be_5 \\
\be_2 &  & \be_6 \\
\be_3 &  & \be_2+\be_4 \\
\end{array}\right\} \leftrightarrow {\color{ProcessBlue} D_1^6}, P=(14)$

\vspace{9pt}

$D_8^6=\left\{ \begin{array}{ccc}
\bzero &  & \be_4 \\
\be_1 &  & \be_5 \\
\be_2 &  & \be_6 \\
\be_3 &  & \be_2+\be_5 \\
\end{array}\right\} \leftrightarrow {\color{ProcessBlue} D_1^6}, P=(15)$

\vspace{9pt}

$D_9^6=\left\{ \begin{array}{ccc}
\bzero &  & \be_4 \\
\be_1 &  & \be_5 \\
\be_2 &  & \be_6 \\
\be_3 &  & \be_2+\be_6 \\
\end{array}\right\} \leftrightarrow {\color{ProcessBlue} D_1^6}, P=(16)$

\vspace{9pt}

$D_{10}^6=\left\{ \begin{array}{ccc}
\bzero &  & \be_4 \\
\be_1 &  & \be_5 \\
\be_2 &  & \be_6 \\
\be_3 &  & \be_3+\be_4 \\
\end{array}\right\} \leftrightarrow {\color{ProcessBlue} D_1^6}, P=(13)(24)$

\vspace{9pt}

$D_{11}^6=\left\{ \begin{array}{ccc}
\bzero &  & \be_4 \\
\be_1 &  & \be_5 \\
\be_2 &  & \be_6 \\
\be_3 &  & \be_3+\be_5 \\
\end{array}\right\} \leftrightarrow {\color{ProcessBlue} D_1^6}, P=(13)(25)$

\vspace{9pt}

$D_{12}^6=\left\{ \begin{array}{ccc}
\bzero &  & \be_4 \\
\be_1 &  & \be_5 \\
\be_2 &  & \be_6 \\
\be_3 &  & \be_3+\be_6 \\
\end{array}\right\} \leftrightarrow {\color{ProcessBlue} D_1^6}, P=(13)(26)$

\vspace{9pt}

$D_{13}^6=\left\{ \begin{array}{ccc}
\bzero &  & \be_4 \\
\be_1 &  & \be_5 \\
\be_2 &  & \be_6 \\
\be_3 &  & \be_4+\be_5 \\
\end{array}\right\} \leftrightarrow {\color{ProcessBlue} D_1^6}, P=(14)(25)$

\vspace{9pt}

$D_{14}^6=\left\{ \begin{array}{ccc}
\bzero &  & \be_4 \\
\be_1 &  & \be_5 \\
\be_2 &  & \be_6 \\
\be_3 &  & \be_4+\be_6 \\
\end{array}\right\} \leftrightarrow {\color{ProcessBlue} D_1^6}, P=(14)(26)$

\vspace{9pt}

$D_{15}^6=\left\{ \begin{array}{ccc}
\bzero &  & \be_4 \\
\be_1 &  & \be_5 \\
\be_2 &  & \be_6 \\
\be_3 &  & \be_5+\be_6 \\
\end{array}\right\} \leftrightarrow {\color{ProcessBlue} D_1^6}, P=(15)(26)$

\vspace{9pt}

{\color{Mulberry}
${\color{Mulberry} D_{16}^6}=\left\{ \begin{array}{cc}
\bzero &   \be_4 \\
\be_1 &   \be_5 \\
\be_2 &   \be_6 \\
\be_3 &   \be_1+\be_2+\be_3 \\
\end{array}\right\}  $}

\vspace{9pt}
$D_{17}^6=\left\{ \begin{array}{cc}
\bzero   & \be_4 \\
\be_1   & \be_5 \\
\be_2   & \be_6 \\
\be_3   & \be_1+\be_3+\be_4 \\
\end{array}\right\} \leftrightarrow {\color{Mulberry} D_{16}^6}, P=(24)$

\vspace{9pt}

$D_{18}^6=\left\{ \begin{array}{cc}
\bzero   & \be_4 \\
\be_1  & \be_5 \\
\be_2  & \be_6 \\
\be_3   & \be_1+\be_4+\be_5 \\
\end{array}\right\} \leftrightarrow {\color{Mulberry} D_{16}^6}, P=(25)(35)$

\vspace{9pt}

$D_{19}^6=\left\{ \begin{array}{cc}
\bzero   & \be_4 \\
\be_1   & \be_5 \\
\be_2   & \be_6 \\
\be_3   & \be_1+\be_5+\be_6 \\
\end{array}\right\} \leftrightarrow {\color{Mulberry} D_{16}^6}, P=(25)(36)$

\vspace{9pt}

$D_{20}^6=\left\{ \begin{array}{cc}
\bzero   & \be_4 \\
\be_1   & \be_5 \\
\be_2   & \be_6 \\
\be_3   & \be_1+\be_2+\be_4 \\
\end{array}\right\} \leftrightarrow {\color{Mulberry} D_{16}^6}, P=(34)$

\vspace{9pt}

$D_{21}^6=\left\{ \begin{array}{cc}
\bzero   & \be_4 \\
\be_1   & \be_5 \\
\be_2   & \be_6 \\
\be_3   & \be_1+\be_2+\be_5 \\
\end{array}\right\} \leftrightarrow {\color{Mulberry} D_{16}^6}, P=(35)$

\vspace{9pt}

$D_{22}^6=\left\{ \begin{array}{cc}
\bzero   & \be_4 \\
\be_1  & \be_5 \\
\be_2   & \be_6 \\
\be_3   & \be_1+\be_2+\be_6 \\
\end{array}\right\} \leftrightarrow {\color{Mulberry} D_{16}^6}, P=(36)$

\vspace{9pt}

$D_{23}^6=\left\{ \begin{array}{cc}
\bzero   & \be_4 \\
\be_1   & \be_5 \\
\be_2   & \be_6 \\
\be_3   & \be_2+\be_3+\be_4 \\
\end{array}\right\} \leftrightarrow {\color{Mulberry} D_{16}^6}, P=(14)$

\vspace{9pt}

$D_{24}^6=\left\{ \begin{array}{cc}
\bzero   & \be_4 \\
\be_1   & \be_5 \\
\be_2   & \be_6 \\
\be_3   & \be_2+\be_4+\be_5 \\
\end{array}\right\} \leftrightarrow {\color{Mulberry} D_{16}^6}, P=(14)(35)$

\vspace{9pt}

$D_{25}^6=\left\{ \begin{array}{cc}
\bzero  & \be_4 \\
\be_1   & \be_5 \\
\be_2   & \be_6 \\
\be_3   & \be_2+\be_5+\be_6 \\
\end{array}\right\} \leftrightarrow {\color{Mulberry} D_{16}^6}, P=(15)(36)$

\vspace{9pt}

$D_{26}^6=\left\{ \begin{array}{cc}
\bzero   & \be_4 \\
\be_1  & \be_5 \\
\be_2   & \be_6 \\
\be_3   & \be_2+\be_3+\be_5 \\
\end{array}\right\} \leftrightarrow {\color{Mulberry} D_{16}^6}, P=(15)$

\vspace{9pt}

$D_{27}^6=\left\{ \begin{array}{cc}
\bzero   & \be_4 \\
\be_1   & \be_5 \\
\be_2   & \be_6 \\
\be_3   & \be_2+\be_3+\be_6 \\
\end{array}\right\} \leftrightarrow {\color{Mulberry} D_{16}^6}, P=(16)$

\vspace{9pt}

$D_{28}^6=\left\{ \begin{array}{cc}
\bzero   & \be_4 \\
\be_1  & \be_5 \\
\be_2  & \be_6 \\
\be_3   & \be_3+\be_4+\be_5 \\
\end{array}\right\} \leftrightarrow {\color{Mulberry} D_{16}^6}, P=(14)(25)$

\vspace{9pt}

$D_{29}^6=\left\{ \begin{array}{cc}
\bzero  & \be_4 \\
\be_1   & \be_5 \\
\be_2   & \be_6 \\
\be_3   & \be_4+\be_5+\be_6 \\
\end{array}\right\} \leftrightarrow {\color{Mulberry} D_{16}^6}, P=(14)(25)(36)$

\vspace{9pt}

$D_{30}^6=\left\{ \begin{array}{cc}
\bzero   & \be_4 \\
\be_1   & \be_5 \\
\be_2   & \be_6 \\
\be_3   & \be_3+\be_5+\be_6 \\
\end{array}\right\} \leftrightarrow {\color{Mulberry} D_{16}^6}, P=(15)(26)$

\vspace{9pt}

$D_{31}^6=\left\{ \begin{array}{cc}
\bzero &   \be_4 \\
\be_1 &   \be_5 \\
\be_2 &   \be_6 \\
\be_3 &   \be_3+\be_4+\be_6 \\
\end{array}\right\} \leftrightarrow {\color{Mulberry} D_{16}^6}, P=(14)(26)$

\vspace{9pt}

$D_{32}^6=\left\{ \begin{array}{cc}
\bzero &   \be_4 \\
\be_1 &   \be_5 \\
\be_2 &   \be_6 \\
\be_3 &   \be_1+\be_3+\be_5 \\
\end{array}\right\} \leftrightarrow {\color{Mulberry} D_{16}^6}, P=(25)$

\vspace{9pt}

$D_{33}^6=\left\{ \begin{array}{cc}
\bzero &   \be_4 \\
\be_1 &   \be_5 \\
\be_2 &   \be_6 \\
\be_3 &   \be_2+\be_4+\be_6 \\
\end{array}\right\} \leftrightarrow {\color{Mulberry} D_{16}^6}, P=(14)(36)$

\vspace{9pt}

$D_{34}^6=\left\{ \begin{array}{cc}
\bzero &   \be_4 \\
\be_1 &   \be_5 \\
\be_2 &   \be_6 \\
\be_3 &   \be_1+\be_3+\be_6 \\
\end{array}\right\} \leftrightarrow {\color{Mulberry} D_{16}^6}, P=(26)$

\vspace{9pt}

$D_{35}^6=\left\{ \begin{array}{cc}
\bzero &   \be_4 \\
\be_1 &   \be_5 \\
\be_2 &   \be_6 \\
\be_3 &   \be_1+\be_4+\be_6 \\
\end{array}\right\} \leftrightarrow {\color{Mulberry} D_{16}^6}, P=(24)(36)$
\vspace{9pt}

$D_{36}^6=\left\{ \begin{array}{cc}
\bzero &   \be_4 \\
\be_1 &  \be_5 \\
\be_2 &  \be_6 \\
\be_3 &   \be_1+\be_2+\be_3+\be_4+\be_5+\be_6 \\
\end{array}\right\}  $

\vspace{9pt}

$\Ran (D)=7$.
\vspace{3pt}

We have $D=\{\bzero, \be_1, \be_2. \be_3,\be_4, \be_5, \be_6,\be_7\}$.

\bibliography{ref.bbl}{}
\bibliographystyle{plain}

\end{document}